\RequirePackage{fix-cm}
\documentclass[natbib,smallextended]{svjour3}       
\usepackage{graphicx}
\usepackage{caption}
\usepackage{subcaption}
\usepackage[flushleft]{threeparttable}
\usepackage{comment}
\usepackage{amsmath}
\usepackage{cases}
\usepackage[utf8]{inputenc}
\usepackage{multirow}
\usepackage{geometry}
\usepackage{float}
 \journalname{Scientometrics}
\begin{document}
\title{EM and $EM^{'}$-index sequence: Construction and application in scientific assessment of scholars
}


\author{Anand Bihari         \and
Sudhakar Tripathi \and
	Akshay Deepak 			\and
	Prabhat Kumar			
}

\institute{Anand Bihari \at
	Department of Computer Science and Engineering\\
	National Institute of Technology Patna, Bihar, India
	\email{anand.cse15@nitp.ac.in, csanandk@gmail.com}          
	\and
	Sudhakar Tripathi \at
	Department of Information Technology\\
	R. E. C. Ambedkar Nagar, Uttar Prasesh, India
	\email{p.stripathi@gmail.com} 
		\and
	Akshay Deepak \at
	Department of Computer Science and Engineering\\
	National Institute of Technology Patna, Bihar, India
	\email{adeepak.cse@nitp.ac.in} 
		\and
	Prabhat Kumar \at
	Department of Computer Science and Engineering\\
	National Institute of Technology Patna, Bihar, India
	\email{prabhat@nitp.ac.in}
}
\date{}

\maketitle

\begin{abstract}
Most of the scientometric indicators use only the total number of citations of an article and produce a single number for scientific assessment of scholars. Although this concept is very simple to compute, it fails to show the scientific productivity and impact of scholars during a time-span or in a year. To overcome this, several time series indicators have been proposed that consider the citations from the entire research career of a scholar. However, these indicators fail to give a comparative assessment of two scholars having same or very similar index value. To overcome this shortcoming, h-index sequence was proposed to assess the impact of scholars during a particular time-span and to compare multiple scholars at a similar stage in their careers. The h-index sequence is based on the h-index formulation. One of the main issues related to the h-index is that it completely ignores the excess citation in scientific assessment; h-index sequence also exhibits a similar behaviour. To overcome these limitations,  in this article, we have discussed the EM and $EM^{'}$-index sequence, and performed an empirical study based on yearly citation count earned from all publications of 89  scholars' publication data. The element of the EM and $EM^{'}$-index sequence for a given year shows the impact of a scholar for that year. We conclude that the EM and $EM^{'}$-index sequence could be used as an alternative metrics to asses the impact of scholars.

\end{abstract}

\section{Introduction}
\label{intro}
The scientometrics and bibliometrics indicators play a key role in the scientific assessment of scholars and are also used in faculty promotion in colleges/universities, scientific award distribution and  project funding etc. \cite{king1987review}. One of the most influential scientometrics/bibliometrics indicator h-index, proposed by \cite{hirsch2005index}, considers productivity as well as impact of scholars. However, this indicator suffers with several drawbacks that restrict its use in comparing scholars having similar index values computed based on their entire research careers. Several studies  have been done to overcome this drawback and increase the acceptability in the scientific assessment (\cite{rousseau2006new,bornmann2008there,rousseau2008reflections,alonso2009h,rosenberg2014biologist,wildgaard2014review}). Generally, most of the indicators produced only a single number to asses the scientific impact of the entire career of the scholars. However, instead of a single number, a set of indicators seem to be more justified in this case.  In this context, first \cite{liang2006h} proposed the h-index sequence. The h-index sequence is the set of h-index values computed from yearly citation counts, where the sequence elements were computed in the reverse chronological order of the academic careers of the scholars, i.e., recent publications being considered first.  However, \cite{egghe2009mathematical} mentioned that the calculation of sequence in the forward direction of time is more precise than the reverse direction, and it is easy to understand as well.  \cite{liu2008definitions} defined 10 different types of time series h-index sequences. \cite{fred2008power} discussed the relationship between the power law model and the h-index sequence using $h_{4}$ values (discussed in Egghe's sequences). \cite{wu2011empirical} performed an empirical study of \textit{real career h-index sequence} based on $h_{3}$ values (discussed in Egghe's sequences).  \cite{liu2014empirical} performed an empirical study of h-index sequence based on yearly citation performance of cumulative publications using $h_{2}$ values obtained from Egghe's sequences. The authors proposed L-sequence obtained from $h_{2}$.   However, all of the above h-index based sequences consider only the $h^{2}$ citation and completely ignore the importance of excess citation. Another issue is that they do not consider all the items in the computation of sequences, whereas the articles that are cited even once have significance in scientific assessment. To overcome this, \cite{Bihari2017} proposed a new measure called EM-index and $EM^{'}$-index. The EM-index gives full credit to highly influential articles, whereas the $EM^{'}$-index considers all articles that are cited even once.

This article proposes the EM and $EM^{'}$-index sequence as an effective way to evaluate the scientific impact of scholars. The EM and $EM^{'}$-index sequence is the sum of the elements calculated  using EM and $EM^{'}$-index formula respectively.
In this article, first we discuss EM-index, $EM^{'}$-index and L-sequence (Sec. \ref{backgroud}). Then we discuss the comparative empirical analysis of EM and $EM^{'}$-index sequence done on yearly citation count earned from all the articles in the dataset of 89 scholars used in \cite{Bihari2018}. The experimental results highlight the properties of EM and $EM^{'}$-index sequence that reflect the overall impact of scholars. Used this way, we show that EM and $EM^{'}$-index sequence provide an alternative superior way to evaluate the scientific impact of scholars.     

\section{Background}\label{backgroud}

The h-index, as proposed in \cite{hirsch2005index}, is described as: ``The h-index of a scholar is $h$ if $h$ of his/her publications have at least $h$ citations each and the rest of the publications may have $h$ or less citations."

This index attracted attention from the research community due to its characteristics, however, it has several limitations. In general, it seems that the most of the indicators give only a single number to show the scientific impact of scholars, but they do not differentiate between scholars having similar index values. Further, they do not take into account the career-duration of scholars. The primary limitation of h-index is that it completely ignores the excess citation (i.e., over and above $h$) of articles. 

\textbf{Example:} Let 15 articles be published by scholar A with the following citation counts Cit=\{30, 30, 25, 22, 22, 21, 15, 15, 14, 10, 10, 10, 9, 8, 1 \}. Let 15 articles be also published by scholar B with the following citation counts cit=\{10, 10, 10, 10, 10, 10, 10, 10, 10, 10, 0, 0, 0, 0, 0\}. In both the cases, the h-index value is 10, however, the scientific impact of scholar A is more than that of scholar B because  scholar A has (i)  more citation counts for each article and (ii)  non-zero citation counts for h-tail articles when compared to scholar B who has zero citation count for h-tail articles.  This shows that h-index may not be a good measure for comparative scientific assessment of scholars  because it  does not give any extra credit to excess citations and h-tail articles. 

To overcome this shortcoming of h-index, the EM-index was proposed by \cite{Bihari2017}; and is defined as: ``The EM-index of an author is the square root of the sum of the elements of the EM-index." Here, the elements of the EM-index of an author are the h-index values computed from the h-core article citation count at multiple levels. The first element of the EM-index is the original h-index and the subsequent elements are the h-index values from the excess citation count of the h-core articles. In the previous example, if we consider the citation count of author A, the components of EM-index are \{10, 6, 5, 3, 2, 2, 2\} and the EM-index is 5.48. Author B has only one  component of EM-index as \{10\} and the EM-index is 3.2. Clearly, the EM-index captures the significant difference in the scientific impact of these two authors. 

The EM-index considers the impact of the excess citation count of the h-core articles, which is not considered in the h-index, but is helpful to differentiate between two different scholars having similar index values. However, the h-index and the EM-index do not consider the impact of all those articles that have been cited even once. If we look at the citation counts of author A, there are some articles  having citation count equal to the h-index or less than that. However,  both the h-index and the EM-index do not consider the impact of citation counts of such articles. To overcome this, a new indicator was  proposed by \cite{Bihari2017} named $EM^{'}$-index. This index is the multidimensional extension of the EM-index.   

In spite of the progress so far, none of the above mentioned indices consider the career-duration of scholars, making it difficult to gauge the impact of a scholar at a particular stage of his/her career. Several articles have been published on this problem; \cite{mahbuba2013year,mahbuba2016new, liu2014empirical} and \cite{Bihari2018} are the recent ones among them. \cite{mahbuba2013year,mahbuba2016new} discussed the year based h-index that considers year wise impact of scholars  based on (i) the total number of citations earned in a particular year, (ii) the total number of citations from all publications that are published in a particular year, and (iii) the total number of publications in a particular year. These year based indices still suffer with h-index limitations. To overcome the year based h-index excess citations problem, \cite{Bihari2018} proposed year based EM and $EM^{'}$-index. The year based indicators cover the entire career of scholars, however they do not consider the year wise impact of scholars. To overcome this issue of year based indices, \cite{liu2014empirical} studied the h-index sequence and proposed a new index called L-sequence. L-sequence considers the entire research career of a scholar to determine the scientific impact. To define L-sequence, consider a scholar who has published $k$ articles in his/her career. Let the first publication year be $y_{1}$ and the current year be $y_{c}$. Then, the L-sequence of an author for $n^{th}$ year, denoted $L_{n}$, is the h-index value computed on the basis of citation counts of all publications received in the $n^{th}$ year. Thus, the L-sequence of the author is  $L_{y_{1}}, L_{y_{1}+1}, ...............L_{y_{c}}$.   

The computation of L-sequence is based on h-index, but it does not account for the impact of excess citations and the h-tail items.

From the above discussion, it can be concluded that the scientific assessment of scholars is done with the help of citations earned by all articles, however, the career-duration of scholars is not considered, which is also significant  in scientific assessment. To overcome this, we propose EM and $EM^{'}$-index sequence, which are described next

\section{EM-index Sequence}
As discussed in the previous section (sec: \ref{backgroud}), the year based indices consider only the total number of citations earned by all publications in a particular year to produce a single number for scientific assessment. In this process, they ignore yearly impact of scholars. Instead of a single number, a set of numbers capturing the yearly impact of scholars will be more suited to evaluate and compare the performance of scholars. To this end, we introduce EM-index sequence based on yearly citations received by all publications.\\
\textbf{Definition of EM-index Sequence}\\
Let the research career of a scholar span $n$ years, publishing $k$ articles. Let $y_{1}$ be the year in which his/her first publication is published. Let the current year be $y_{c}$. The  EM-index sequence element for the $y^{th}$ year, denoted $EM_y$, is calculated from the citation count in $y^{th}$ year using the EM-index formula given  in \cite{Bihari2017}. Then, the EM-index sequence value is computed as the sum of all such EM-index sequence elements. Formally,  
\begin{equation}
EM-index~ Sequence=\sum_{y=1}^{c} EM_{y}
\end{equation} 
For example, consider the scientific research history and impact of author \textbf{Andrew D. Jackson} (Source: Web of Science) as shown in Table \ref{tab:Publication history and year wise impact of Author Andrew D. Jackson}. The corresponding EM-index sequence elements  are 2.24, 2.65, 3.16, 2.83, 2.24, 3.16, 3.74, 3, 3.32, 2.45 \& 1.73 and the EM-index sequence value is 29.17. 
\begin{table}[hbt]
	\renewcommand{\arraystretch}{1.2}
	\caption{Publication history and year wise impact of Author Andrew D. Jackson}
	\label{tab:Publication history and year wise impact of Author Andrew D. Jackson}
	\centering
	\begin{tabular}{p{1.8cm}p{0.8cm}p{0.8cm}p{0.8cm}p{0.8cm}p{0.8cm}p{0.8cm}p{0.8cm}p{0.8cm}p{0.8cm}p{0.8cm}p{0.8cm}}		\hline
\textbf{Publication Year $\downarrow$}	&	\textbf{2007}	&	\textbf{2008}	&	\textbf{2009}	&	\textbf{2010}	&	\textbf{2011}	&	\textbf{2012}	&	\textbf{2013}	&	\textbf{2014}	&	\textbf{2015}	&	\textbf{2016}	&	\textbf{2017}	\\\hline
2006	&	11	&	9	&	11	&	15	&	12	&	11	&	20	&	14	&	16	&	11	&	5	\\
2007	&	4	&	6	&	9	&	8	&	5	&	10	&	13	&	8	&	4	&	2	&	0	\\
2007	&	0	&	0	&	0	&	0	&	0	&	0	&	0	&	0	&	0	&	0	&	0	\\
2008	&	0	&	1	&	10	&	4	&	4	&	5	&	7	&	6	&	4	&	5	&	1	\\
2008	&	0	&	0	&	0	&	0	&	0	&	1	&	0	&	0	&	0	&	0	&	0	\\
2008	&	0	&	0	&	0	&	0	&	0	&	0	&	0	&	0	&	0	&	0	&	0	\\
2009	&	0	&	0	&	0	&	3	&	5	&	9	&	4	&	6	&	10	&	6	&	2	\\
2009	&	0	&	0	&	0	&	0	&	0	&	2	&	0	&	1	&	0	&	0	&	0	\\
2010	&	0	&	0	&	0	&	0	&	1	&	0	&	0	&	0	&	0	&	0	&	0	\\
2011	&	0	&	0	&	0	&	0	&	1	&	3	&	3	&	4	&	2	&	1	&	0	\\
2013	&	0	&	0	&	0	&	0	&	0	&	0	&	0	&	1	&	1	&	1	&	1	\\
2013	&	0	&	0	&	0	&	0	&	0	&	0	&	0	&	0	&	1	&	0	&	0	\\
2013	&	0	&	0	&	0	&	0	&	0	&	0	&	0	&	0	&	0	&	0	&	0	\\
2013	&	0	&	0	&	0	&	0	&	0	&	0	&	0	&	0	&	0	&	0	&	0	\\
2014	&	0	&	0	&	0	&	0	&	0	&	0	&	0	&	0	&	0	&	0	&	0	\\
2015	&	0	&	0	&	0	&	0	&	0	&	0	&	0	&	0	&	1	&	1	&	0	\\
2015	&	0	&	0	&	0	&	0	&	0	&	0	&	0	&	0	&	0	&	0	&	0	\\
2015	&	0	&	0	&	0	&	0	&	0	&	0	&	0	&	0	&	0	&	0	&	0	\\
2015	&	0	&	0	&	0	&	0	&	0	&	0	&	0	&	0	&	0	&	0	&	0	\\
2015	&	0	&	0	&	0	&	0	&	0	&	0	&	0	&	0	&	0	&	0	&	0	\\
2015	&	0	&	0	&	0	&	0	&	0	&	0	&	0	&	0	&	0	&	0	&	0	\\
2016	&	0	&	0	&	0	&	0	&	0	&	0	&	0	&	0	&	0	&	1	&	0	\\
2016	&	0	&	0	&	0	&	0	&	0	&	0	&	0	&	0	&	0	&	0	&	0	\\
2016	&	0	&	0	&	0	&	0	&	0	&	0	&	0	&	0	&	0	&	0	&	0	\\
2016	&	0	&	0	&	0	&	0	&	0	&	0	&	0	&	0	&	0	&	0	&	0	\\
2017	&	0	&	0	&	0	&	0	&	0	&	0	&	0	&	0	&	0	&	0	&	0	\\
EM-index &2.24 &2.65 &3.16 &2.83 &2.24 &3.16 &3.74 &3 &3.32 &2.45 &1.73\\
\hline
\end{tabular}
\end{table}
\section{Empirical  Study of EM-index-sequence} \label{Empirical  Study of EM-index-sequence}
In this section, we present an empirical study of EM-index sequence over the h-index sequence, h-index, EM-index, $EM^{'}$-index  and year based EM-index by citations. The study illustrates the merit of EM-index sequence in comparison to others. To do this, we have used the publication-data of 89 scholars from reference \cite{Bihari2018}. Among the 89 scholars, most of them are working in scientometric and biblometrics fields of research. The h-index, EM-index, $EM^{'}$-index, year based EM-index by citations, h-index sequence and EM-index sequence for all 89 scholars is shown in Figure \ref{fig:h-index, EM-index, $EM^{'}$-index, h-index sequences and EM-index sequences}. As can be seen, there is significant variation in the indices values of all scholars. Table \ref{tab:Result of EM-index sequences} shows the corresponding EM-index sequence value for each of the 89 scholars. 
\begin{figure}[H]
\begin{subfigure}{.5\textwidth}
	\centering
	\includegraphics[width=2.8in,height=2in,keepaspectratio]{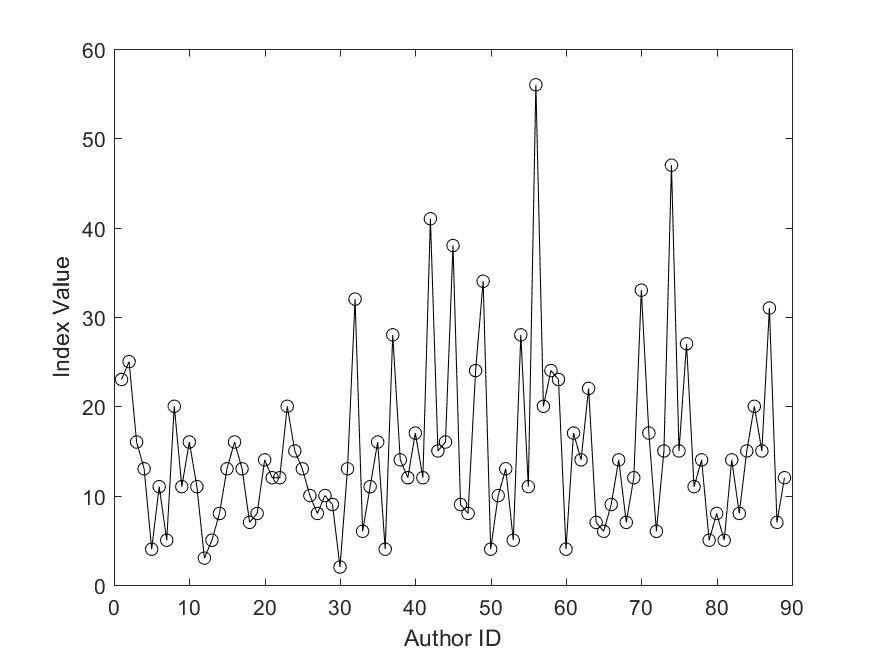}
	\caption{The h-index of all Scholars}
	\label{fig:a}
\end{subfigure}%
\begin{subfigure}{.5\textwidth}
	\centering
	\includegraphics[width=2.8in,height=2in,keepaspectratio]{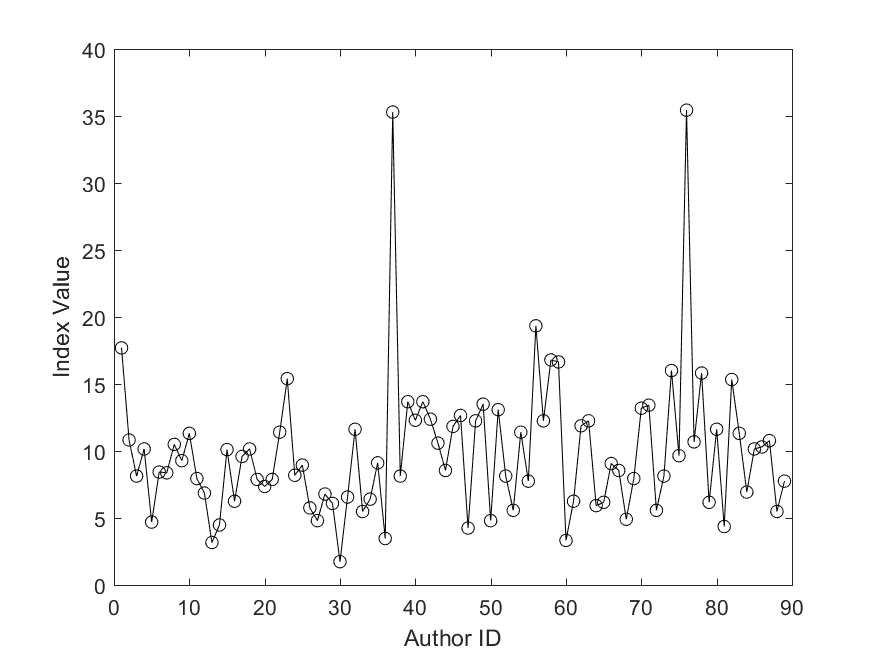}
	\caption{The EM-index of all Scholars}
	\label{fig:b}
\end{subfigure}
\end{figure}
\begin{figure}
\setcounter{figure}{0}
\begin{subfigure}{.5\textwidth}
	\setcounter{subfigure}{2}
	\centering
	\includegraphics[width=2.8in,height=2in,keepaspectratio]{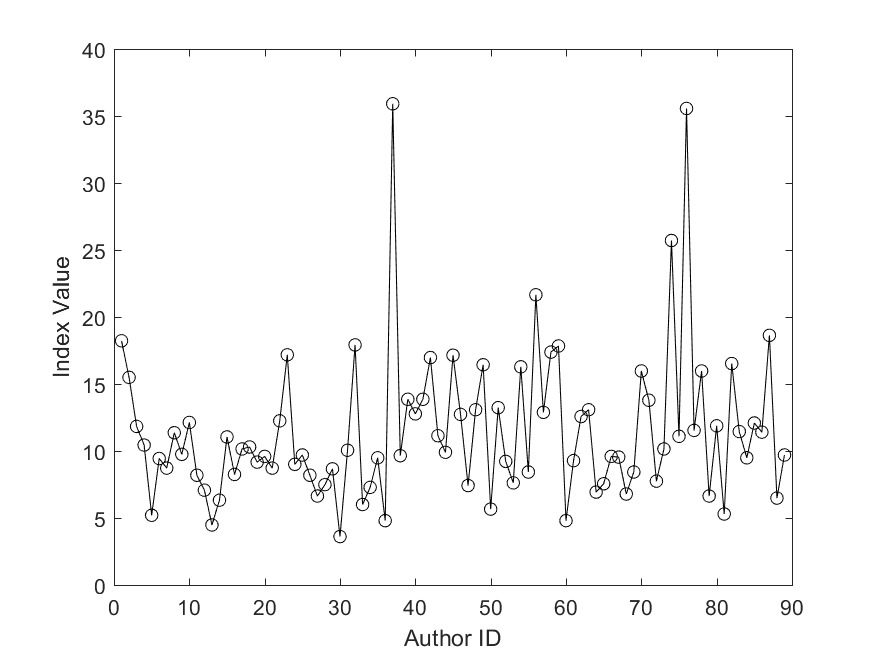}
	\caption{The $EM^{'}$-index of all Scholars}
	\label{fig:b}
\end{subfigure}
\begin{subfigure}{.5\textwidth}
	\centering
	\includegraphics[width=2.8in,height=2in,keepaspectratio]{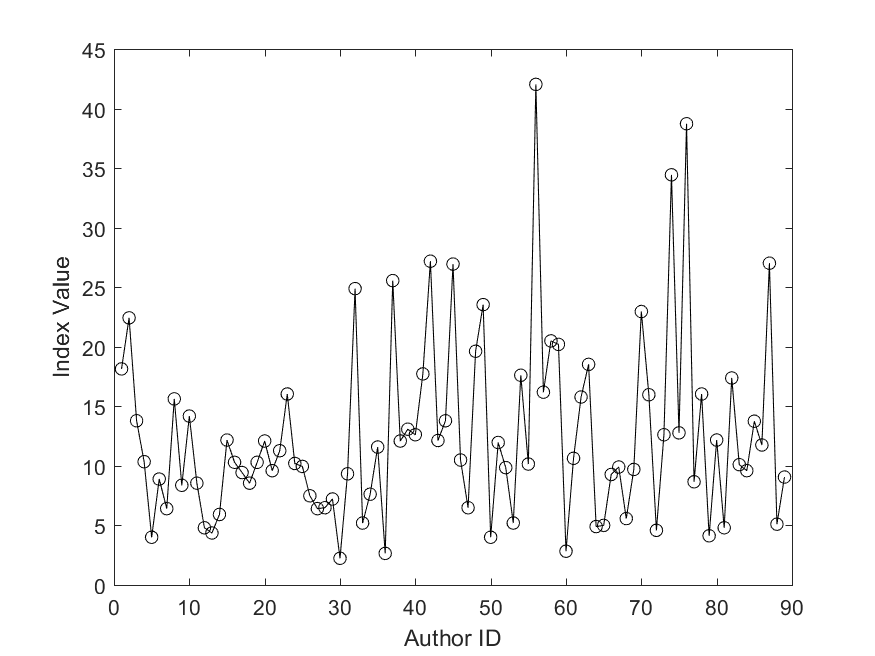}
	\caption{The Year based EM-index by citations of all Scholars}
	\label{fig:b}
\end{subfigure}
\begin{subfigure}{.5\textwidth}
	\setcounter{subfigure}{2}
	\centering
	\includegraphics[width=2.8in,height=2in,keepaspectratio]{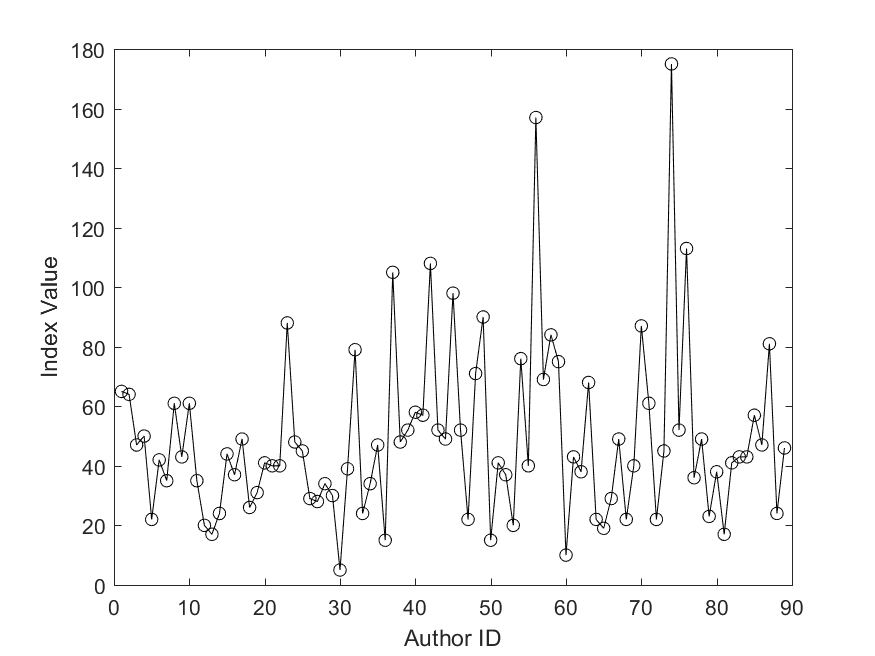}
	\caption{The h-index sequence of all Scholars}
	\label{fig:b}
\end{subfigure}
\begin{subfigure}{.5\textwidth}
	\centering
	\includegraphics[width=2.8in,height=2in,keepaspectratio]{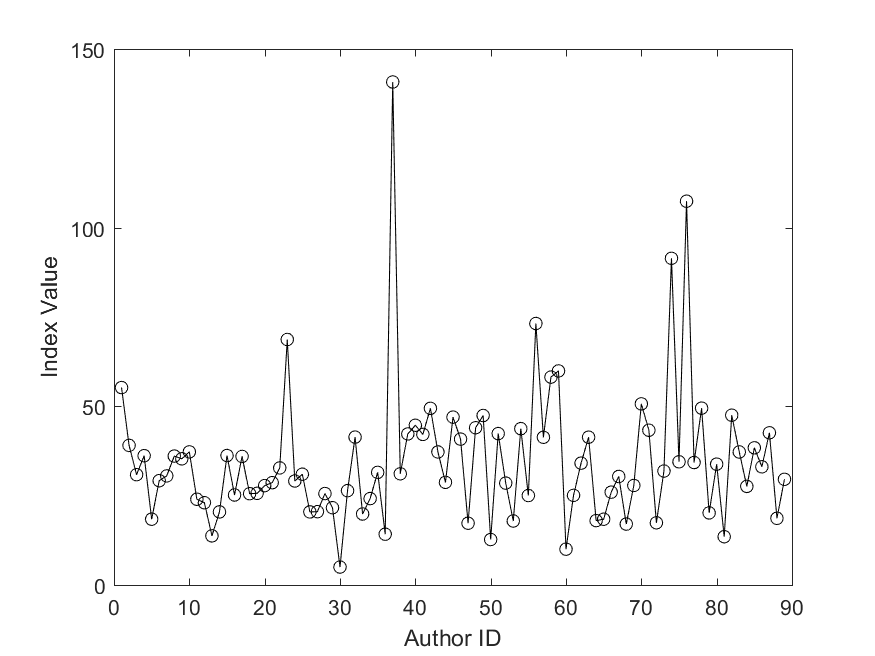}
	\caption{The EM-index sequence of all Scholars}
	\label{fig:b}
\end{subfigure}
\caption{h-index, EM-index, $EM^{'}$-index, year based EM-index by citations, h-index sequence and EM-index sequence for the publication-data of 89 scholars from reference \cite{Bihari2018}}
\label{fig:h-index, EM-index, $EM^{'}$-index, h-index sequences and EM-index sequences}
\end{figure}

\inputencoding{utf8}
\begin{table}[h!]
	\renewcommand{\arraystretch}{1.3}
	\caption{The EM-index sequence of all 89 scholars}
	\label{tab:Result of EM-index sequences}
	\centering
	\begin{tabular}{clp{2cm}clp{2cm}}		\hline
		\textbf{ID}     &\textbf{Author} &\textbf{EM-index Sequence}&\textbf{ID}     &\textbf{Author} &\textbf{EM-index Sequence}  \\\hline
1	&	Adamantios Diamantopoulos	&	55.24	&	46	&	Lokman Meho	&	40.81	\\
2	&	Albert Zomaya	&	39.06	&	47	&	Luca  Mastrogiacomo	&	17.26	\\
3	&	Alireza Abbasi	&	30.85	&	48	&	Ludo Waltman	&	44.00	\\
4	&	András Schubert	&	36.16	&	49	&	Lutz bornmann	&	47.42	\\
5	&	András Telcs	&	18.39	&	50	&	Maisano, DA	&	12.70	\\
6	&	Andreas Thor	&	29.17	&	51	&	Marek Kosmulski	&	42.35	\\
7	&	Andrew D. Jackson	&	30.51	&	52	&	Maria Bordons	&	28.49	\\
8	&	Anne-Wil	&	36.07	&	53	&	Mark Fine	&	17.89	\\
9	&	Aric Hagberg	&	35.29	&	54	&	Mark Newman	&	43.74	\\
10	&	Barry Bozeman	&	37.23	&	55	&	Mathieu Ouimet	&	25.01	\\
11	&	Ben R Martin	&	23.98	&	56	&	Matjaž Perc	&	73.17	\\
12	&	Benny Lautrup	&	23.00	&	57	&	Matthew O. Jackson	&	41.33	\\
13	&	Berwin Turlach	&	13.73	&	58	&	Mauno Vihinen	&	58.17	\\
14	&	Birger Larsen	&	20.44	&	59	&	Michael Jennions	&	59.89	\\
15	&	Blaise Cronin	&	36.21	&	60	&	Michael L. Nelson	&	10.00	\\
16	&	C Lee Giles	&	25.17	&	61	&	Miguel A.	&	25.07	\\
17	&	Carlos Pecharroman	&	35.97	&	62	&	Morten Schmidt	&	34.08	\\
18	&	Caroline S. Wagner	&	25.53	&	63	&	Nees Jan van Eck	&	41.35	\\
	\hline
\end{tabular}
\end{table}
\begin{table}[]
\addtocounter{table}{-1}
\renewcommand{\arraystretch}{1.2}
	\caption{The EM-index sequence of all 89 scholars continued......}
\centering
\begin{tabular}{clp{2cm}clp{2cm}}		\hline
	\textbf{ID}     &\textbf{Author} &\textbf{EM-index Sequence}&\textbf{ID}     &\textbf{Author} &\textbf{EM-index Sequence}  \\\hline

19	&	Christoph Bartneck	&	25.64	&	64	&	Nils T. Hagen	&	17.99	\\
20	&	Claes Wohlin	&	27.79	&	65	&	Olle Persson	&	18.39	\\
21	&	Clint D. Kelly	&	28.62	&	66	&	Paul Wouters	&	25.96	\\
22	&	Dimitrios Katsaros	&	32.75	&	67	&	Peter Jacso	&	30.36	\\
23	&	Egghe Leo	&	68.69	&	68	&	Raf Guns	&	17.02	\\
24	&	Elizabeth A. Corley	&	29.09	&	69	&	Raj Kumar Pan	&	27.83	\\
25	&	Erhard Rahm	&	30.98	&	70	&	Richard S J Tol	&	50.67	\\
26	&	Fiorenzo  Franceschini	&	20.43	&	71	&	Roberto Todeschini	&	43.30	\\
27	&	Fred Y.	&	20.50	&	72	&	Robin hankin	&	17.39	\\
28	&	Gad Saad	&	25.56	&	73	&	Rodrigo Costas	&	31.87	\\
29	&	Gangan Prathap	&	21.60	&	74	&	Ronald Rouseau	&	91.43	\\
30	&	Gary M. Olson	&	5.00	&	75	&	Ruediger mutz	&	34.46	\\
31	&	Gerhard Woeginger	&	26.38	&	76	&	Santo Fortunato	&	107.41	\\
32	&	Guang-Hong Yang	&	41.37	&	77	&	Serge GALAM	&	34.26	\\
33	&	Heidi Winklhofer	&	19.85	&	78	&	Sergio  Alonso	&	49.48	\\
34	&	Hendrik P. van Dalen	&	24.16	&	79	&	Steve Lawrence	&	20.16	\\
35	&	Henk F. Moed	&	31.51	&	80	&	Sune Lehmann	&	33.80	\\
36	&	Herbert Van de Sompel	&	14.20	&	81	&	Terttu Luukkonen	&	13.49	\\
37	&	Hirsche	&	140.79	&	82	&	Vicenç	&	47.50	\\
38	&	James Moody	&	31.10	&	83	&	Walter W (Woody) Powell	&	37.21	\\
39	&	Jayant Vaidya	&	42.25	&	84	&	Werner Marx	&	27.58	\\
40	&	Jerome Vanclay	&	44.69	&	85	&	Wolfgang Glänzel	&	38.37	\\
41	&	Johan Bollen	&	42.13	&	86	&	Yannis Manolopoulos	&	33.06	\\
42	&	JOHN IRVINE	&	49.43	&	87	&	Ying Ding	&	42.58	\\
43	&	Judit Bar-Ilan	&	37.21	&	88	&	Yu-Hsin Liu	&	18.65	\\
44	&	Kène Henkens	&	28.71	&	89	&	Yvonne Rogers	&	29.57	\\
45	&	Loet Leydesdorff	&	46.97	&		&		&		\\

\hline
\end{tabular}
\end{table}

The EM-index sequence provides a more balanced and efficient way to assess the scientific impact of scholars. The EM-index sequence elements provide year wise scientific impact of a scholar that helps to compare the performance of scholars at a particular year of their career. In order to validate this, a comparative analysis has been made with h-index sequence as shown in Table \ref{tab:A Comparative Result of h-index and EM-index sequences}. 
\inputencoding{utf8}
\begin{table}[h!]
	\renewcommand{\arraystretch}{1.3}
	\caption{Comparison of h-index sequence with EM-index sequence}
	\label{tab:A Comparative Result of h-index and EM-index sequences}
	\centering
	\begin{tabular}{clp{2cm}clp{2cm}}		\hline
		\textbf{ID}     &\textbf{Author} &\textbf{h-index Sequence}&\textbf{Rank}     &\textbf{EM-index Sequence} &\textbf{Rank}  \\\hline
		1	&	Adamantios Diamantopoulos	&	65	&	18	&	55.24	&	8	\\
		2	&	Albert Zomaya	&	64	&	19	&	39.06	&	27	\\
		3	&	Alireza Abbasi	&	47	&	37	&	30.85	&	47	\\
		4	&	András Schubert	&	50	&	30	&	36.16	&	33	\\
		5	&	András Telcs	&	22	&	76	&	18.39	&	77	\\
		6	&	Andreas Thor	&	42	&	48	&	29.17	&	51	\\
		7	&	Andrew D. Jackson	&	35	&	62	&	30.51	&	48	\\
		8	&	Anne-Wil Harzing	&	61	&	20	&	36.07	&	34	\\
		9	&	Aric Hagberg	&	43	&	44	&	35.29	&	36	\\
		10	&	Barry Bozeman	&	61	&	21	&	37.23	&	29	\\
		11	&	Ben R Martin	&	35	&	63	&	23.98	&	68	\\
		12	&	Benny Lautrup	&	20	&	81	&	23.00	&	69	\\
			\hline
	\end{tabular}
\end{table}
\begin{table}[h!]
\addtocounter{table}{-1}
\renewcommand{\arraystretch}{1.2}
\caption{Comparison of h-index sequence with EM-index sequence continued....}
\centering
\begin{tabular}{clp{2cm}clp{2cm}}		\hline
	\textbf{ID}     &\textbf{Author} &\textbf{h-index Sequence}&\textbf{Rank}     &\textbf{EM-index Sequence} &\textbf{Rank}  \\\hline
	13	&	Berwin Turlach	&	17	&	84	&	13.73	&	85	\\
	14	&	Birger Larsen	&	24	&	72	&	20.44	&	72	\\
	15	&	Blaise Cronin	&	44	&	43	&	36.21	&	32	\\
	16	&	C Lee Giles	&	37	&	59	&	25.17	&	64	\\
	17	&	Carlos Pecharroman	&	49	&	31	&	35.97	&	35	\\
	18	&	Caroline S. Wagner	&	26	&	71	&	25.53	&	63	\\
	19	&	Christoph Bartneck	&	31	&	66	&	25.64	&	61	\\
	20	&	Claes Wohlin	&	41	&	49	&	27.79	&	57	\\
	21	&	Clint D. Kelly	&	40	&	52	&	28.62	&	54	\\
	22	&	Dimitrios Katsaros	&	40	&	53	&	32.75	&	42	\\
	23	&	Egghe Leo	&	88	&	8	&	68.69	&	5	\\
	24	&	Elizabeth A. Corley	&	48	&	35	&	29.09	&	52	\\
	25	&	Erhard Rahm	&	45	&	41	&	30.98	&	46	\\
	26	&	Fiorenzo  Franceschini	&	29	&	68	&	20.43	&	73	\\
	27	&	Fred Y. Ye	&	28	&	70	&	20.50	&	71	\\
	28	&	Gad Saad	&	34	&	64	&	25.56	&	62	\\
	29	&	Gangan Prathap	&	30	&	67	&	21.60	&	70	\\
	30	&	Gary M. Olson	&	5	&	89	&	5.00	&	89	\\
	31	&	Gerhard Woeginger	&	39	&	56	&	26.38	&	59	\\
	32	&	Guang-Hong Yang	&	79	&	12	&	41.37	&	23	\\
	33	&	Heidi Winklhofer	&	24	&	73	&	19.85	&	75	\\
	34	&	Hendrik P. van Dalen	&	34	&	65	&	24.16	&	67	\\
	35	&	Henk F. Moed	&	47	&	38	&	31.51	&	44	\\
	36	&	Herbert Van de Sompel	&	15	&	86	&	14.20	&	84	\\
	37	&	Hirsch J.E.	&	105	&	5	&	140.79	&	1	\\
	38	&	James Moody	&	48	&	36	&	31.10	&	45	\\
	39	&	Jayant Vaidya	&	52	&	26	&	42.25	&	21	\\
	40	&	Jerome Vanclay	&	58	&	23	&	44.69	&	15	\\
	41	&	Johan Bollen	&	57	&	24	&	42.13	&	22	\\
	42	&	John Irvine	&	108	&	4	&	49.43	&	11	\\
	43	&	Judit Bar-Ilan	&	52	&	27	&	37.21	&	30	\\
		44	&	Kène Henkens	&	49	&	32	&	28.71	&	53	\\
		45	&	Loet Leydesdorff	&	98	&	6	&	46.97	&	14	\\
		46	&	Lokman Meho	&	52	&	28	&	40.81	&	26	\\
		47	&	Luca  Mastrogiacomo	&	22	&	77	&	17.26	&	82	\\
		48	&	Ludo Waltman	&	71	&	15	&	44.00	&	16	\\
		49	&	Lutz Bornmann	&	90	&	7	&	47.42	&	13	\\
		50	&	Maisano, Domenico A.	&	15	&	87	&	12.70	&	87	\\
		51	&	Marek Kosmulski	&	41	&	50	&	42.35	&	20	\\
		52	&	Maria Bordons	&	37	&	60	&	28.49	&	55	\\
		53	&	Mark Fine	&	20	&	82	&	17.89	&	80	\\
		54	&	Mark Newman	&	76	&	13	&	43.74	&	17	\\
		55	&	Mathieu Ouimet	&	40	&	54	&	25.01	&	66	\\
		56	&	Matjaž Perc	&	157	&	2	&	73.17	&	4	\\
		57	&	Matthew O. Jackson	&	69	&	16	&	41.33	&	25	\\
		58	&	Mauno Vihinen	&	84	&	10	&	58.17	&	7	\\
		59	&	Michael Jennions	&	75	&	14	&	59.89	&	6	\\
		60	&	Michael L. Nelson	&	10	&	88	&	10.00	&	88	\\
		61	&	Miguel A. García-Pérez	&	43	&	45	&	25.07	&	65	\\
		62	&	Morten Schmidt	&	38	&	57	&	34.08	&	39	\\
		63	&	Nees Jan van Eck	&	68	&	17	&	41.35	&	24	\\
		64	&	Nils T. Hagen	&	22	&	78	&	17.99	&	79	\\
		65	&	Olle Persson	&	19	&	83	&	18.39	&	78	\\	
		
		\hline
	\end{tabular}
\end{table}
\begin{table}[h!]
\addtocounter{table}{-1}
\renewcommand{\arraystretch}{1.2}
\caption{Comparison of h-index sequence with EM-index sequence continued....}
\centering
\begin{tabular}{clp{2cm}clp{2cm}}		\hline
\textbf{ID}     &\textbf{Author} &\textbf{h-index Sequence}&\textbf{Rank}     &\textbf{EM-index Sequence} &\textbf{Rank}  \\\hline
	66	&	Paul Wouters	&	29	&	69	&	25.96	&	60	\\
	67	&	Peter Jacso	&	49	&	33	&	30.36	&	49	\\
		68	&	Raf Guns	&	22	&	79	&	17.02	&	83	\\
		69	&	Raj Kumar Pan	&	40	&	55	&	27.83	&	56	\\
		70	&	Richard S. J. Tol	&	87	&	9	&	50.67	&	9	\\
		71	&	Roberto Todeschini	&	61	&	22	&	43.30	&	18	\\
		72	&	Robin Hankin	&	22	&	80	&	17.39	&	81	\\
		73	&	Rodrigo Costas	&	45	&	42	&	31.87	&	43	\\
		74	&	Ronald Rousseau	&	175	&	1	&	91.43	&	3	\\
		75	&	Ruediger mutz	&	52	&	29	&	34.46	&	37	\\
		76	&	Santo Fortunato	&	113	&	3	&	107.41	&	2	\\
		77	&	Serge Galam	&	36	&	61	&	34.26	&	38	\\
		78	&	Sergio  Alonso	&	49	&	34	&	49.48	&	10	\\
		79	&	Steve Lawrence	&	23	&	75	&	20.16	&	74	\\
		80	&	Sune Lehmann	&	38	&	58	&	33.80	&	40	\\
		81	&	Terttu Luukkonen	&	17	&	85	&	13.49	&	86	\\
		82	&	Vicenç	&	41	&	51	&	47.50	&	12	\\
		83	&	Walter W (Woody) Powell	&	43	&	46	&	37.21	&	31	\\
		84	&	Werner Marx	&	43	&	47	&	27.58	&	58	\\
		85	&	Wolfgang Glänzel	&	57	&	25	&	38.37	&	28	\\
		86	&	Yannis Manolopoulos	&	47	&	39	&	33.06	&	41	\\
		87	&	Ying Ding	&	81	&	11	&	42.58	&	19	\\
		88	&	Yu-Hsin Liu	&	24	&	74	&	18.65	&	76	\\
		89	&	Yvonne Rogers	&	46	&	40	&	29.57	&	50	\\
		
		\hline
	\end{tabular}
\end{table}

The comparative result of h-index sequence and EM-index sequence values shows that the excess citation of articles can increase the impact of scholars, improving their rank in the group. For example:  
\begin{enumerate}
	\item Author \textbf{Hirsch J.E.} (Author ID=37) is ranked 5 as per the corresponding h-index sequence value of 105. His rank improves to 1 when computed as per the corresponding EM-index sequence value of 140.79. This is because the h-index sequence did not account for 4122 excess citations that helped in improving the rank computed based on EM-index sequence. 
	\item On the other hand, Author \textbf{Ronald Rousseau} (Author ID=74) is ranked 1 as per the corresponding h-index sequence value of 175. His rank reduces to 3 when computed as per the corresponding EM-index sequence value of 91.43. This is because the excess citation count of Author \textbf{Ronald Rousseau}, which is 1774, does not match up to the very high h-index sequence value of 175, resulting in a little bit lowering of rank computed based on EM-index sequence.
	\item As an extreme example, author \textbf{Marek Kosmulski} (Author ID=51) is ranked 50 based on h-index sequence value of 41, whereas his rank is 20 based on EM-index sequence value of 42.35. Here, a marginal difference in the two index values affects the corresponding ranks in a major way. Here, the author has 435 excess citations.
\end{enumerate}

\begin{figure}[H]
	\begin{subfigure}{.48\textwidth}
		\centering
		\includegraphics[width=2.8in,height=2in,keepaspectratio]{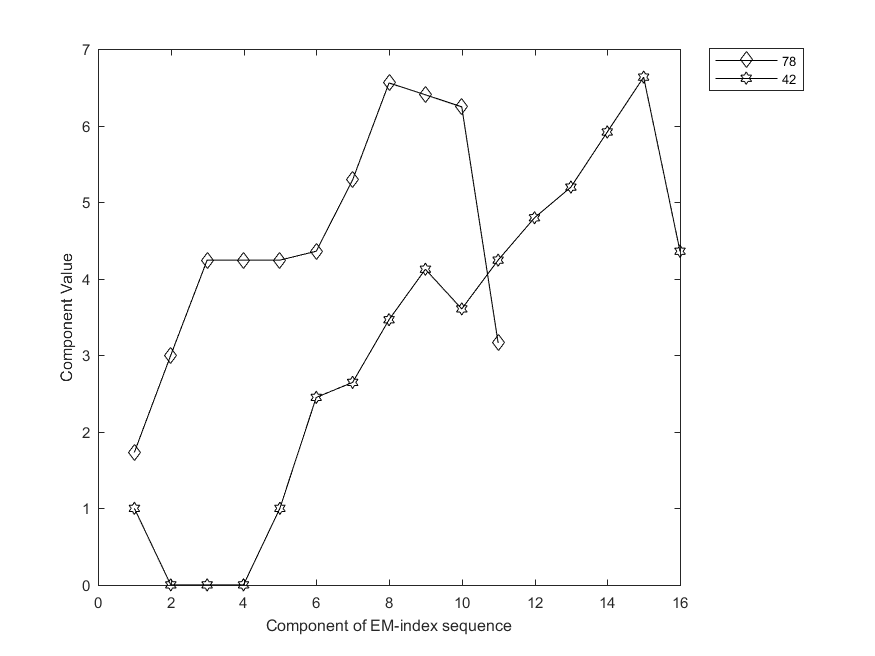}
		\caption{The component analysis of EM-index sequence ~~~~~~~~~~~~of Author ID- 78 \& 42 with corresponding index value 49.48 \& 49.43}
		\label{fig:a}
	\end{subfigure}
	\begin{subfigure}{.5\textwidth}
		\centering
		\includegraphics[width=2.8in,height=2in,keepaspectratio]{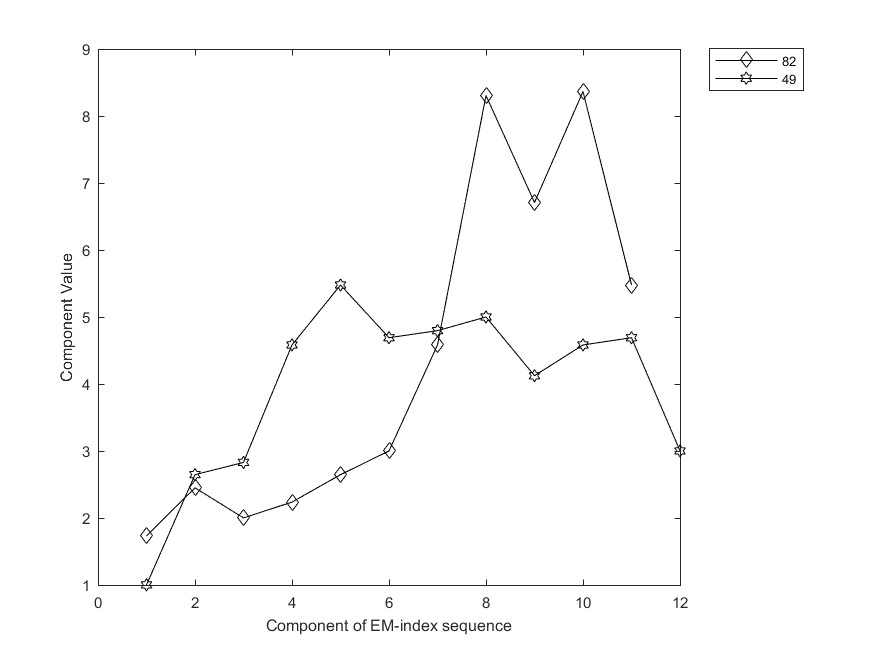}
		\caption{The component analysis of EM-index sequence ~~~~~~~~~~~~ of Author ID- 82 \& 49 with corresponding index value 47.50 \& 47.42}
		\label{fig:b}
	\end{subfigure}
	\begin{subfigure}{.5\textwidth}
		\centering
		\includegraphics[width=2.8in,height=2in,keepaspectratio]{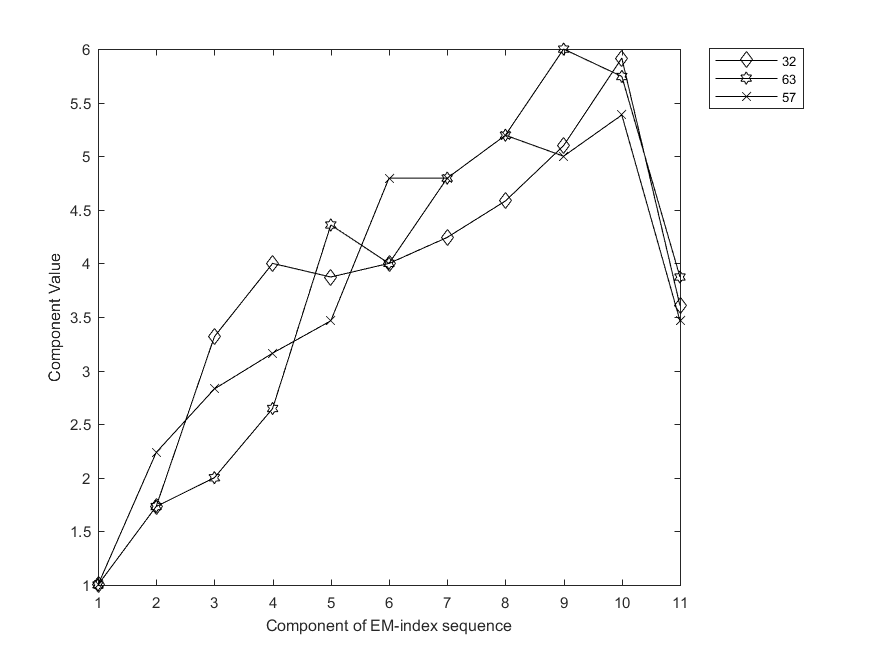}
		\caption{The component analysis of EM-index sequence of Author ID- 32,63 \& 57 with corresponding index value 41.37, 41.35 \& 41.33}
		\label{fig:b}
	\end{subfigure}
	\caption{The component analysis of $EM$-index sequence}
	\label{fig:The component analysis of $EM$-index sequence}
\end{figure}

In general, the comparative analysis between scholars has been performed based on total index value. In this case, if two scholars get the same index value, then it is very difficult to discriminate the impact of scholars. Hence, instead of a single number, a set of numbers are more suited to compare the scientific impact of two  scholars.   Consider one such case: Figure \ref{fig:The component analysis of $EM$-index sequence} shows the EM-index sequence component analysis of scholars, where all have very similar EM-index sequence values.  As shown in Figure \ref{fig:The component analysis of $EM$-index sequence}(a), authors 78 and  42 have EM-index sequence values of  49.48 and 49.43 respectively; their career spans 12 and 18 years respectively. Even though both authors have similar EM-index sequence values, the scientific impact of author 78 is more than that of author 42 considering the latter's career so far, i.e., the first 12 years only. Figure \ref{fig:The component analysis of $EM$-index sequence}(b) compares scholars 82 (EM-index sequence value: 47.50, career span: 13 years) and 49 (EM-index sequence value: 47.42, career span: 12 years). Here both authors have similar career spans, however, as evident from the plot, author 49 has a greater scientific impact during the first half of his career, i.e., the first seven years, while scholar 82 has a greater scientific impact during the second half, i.e., the last four years. This is not evident by just looking at the EM-index sequence values of the two scholars. Figure \ref{fig:The component analysis of $EM$-index sequence}(c) compares scholars 32 (EM-index sequence value: 41.37, career span: 12 years), 63 (EM-index sequence value: 41.35, career span: 12 years) and 57 (EM-index sequence value: 41.33, career span: 12 years). Here all scholars have similar career spans and their impacts are almost the  same.


From the above discussion, it is clear that the elements of  EM-index sequence helps in comparing the scientific impacts of scholars having equal or very similar EM-index sequence values.

While EM-index sequence values are better than h-index sequence values with respect to the excess citations of h-core articles, both can be limited due to complete ignorance of the citations of h-tail articles. This is not good because some of the h-tail articles have citation count similar to the h-core articles. Such h-tail articles are obviously important, however, even the h-tail articles with lesser citations can be significant in assessing the impact of a scholar, and hence, should not be ignored. Motivated by this, reference \cite{garcia2009multidimensional} proposed multidimensional h-index and \cite{Bihari2017} proposed $EM^{'}$-index. Both the proposed indices consider all publications with non-zero citations. Next, we introduce $EM^{'}$-index sequence and discuss how it can be used  (i) to better assess the scientific impact of a scholar, and (ii) to compare scholars with similar career spans or having equal number of publications.  

\section{$EM^{'}$-index sequence}


Let the research career of a scholar span $n$ years publishing $k$ articles. Let $y_{1}$ be the first year of publications and $y_{c}$ be the current year. The $EM^{'}$-index sequence element for the $y^{th}$ year, denoted $EM^{'}_{y}$, is computed from the citation count in the $y^{th}$ year using $EM^{'}$-index formula as given in \cite{Bihari2017}). Then, the $EM^{'}$-index sequence is the sum of all such $EM^{'}$-index sequence elements. Formally, it is defined as: 
\begin{equation}
EM^{'}-index~ Sequences=\sum_{y=1}^{c} EM^{'}_{y}
\end{equation} 

To demonstrate the effectiveness of $EM^{'}$-index sequence, consider the data of author \textbf{Andrew D. Jackson} from Table \ref{tab:Publication history and year wise impact of Author Andrew D. Jackson}. The corresponding $EM^{'}$-index sequence values are shown in Table \ref{tab:Publication history and year wise impact of Author Andrew D. Jacksonfor }.

\begin{table}[hbt]
	\renewcommand{\arraystretch}{1.2}
	\caption{Publication history and year wise impact of Author Andrew D. Jackson for $EM^{'}$-index sequence}
	\label{tab:Publication history and year wise impact of Author Andrew D. Jacksonfor }
	\centering
	\begin{tabular}{p{1.8cm}p{0.8cm}p{0.8cm}p{0.8cm}p{0.8cm}p{0.8cm}p{0.8cm}p{0.8cm}p{0.8cm}p{0.8cm}p{0.8cm}p{0.8cm}}		\hline
		\textbf{Publication Year}	&	\textbf{2007}	&	\textbf{2008}	&	\textbf{2009}	&	\textbf{2010}	&	\textbf{2011}	&	\textbf{2012}	&	\textbf{2013}	&	\textbf{2014}	&	\textbf{2015}	&	\textbf{2016}	&	\textbf{2017}	\\\hline
		2006	&	11	&	9	&	11	&	15	&	12	&	11	&	20	&	14	&	16	&	11	&	5	\\
		2007	&	4	&	6	&	9	&	8	&	5	&	10	&	13	&	8	&	4	&	2	&	0	\\
		2007	&	0	&	0	&	0	&	0	&	0	&	0	&	0	&	0	&	0	&	0	&	0	\\
		2008	&	0	&	1	&	10	&	4	&	4	&	5	&	7	&	6	&	4	&	5	&	1	\\
		2008	&	0	&	0	&	0	&	0	&	0	&	1	&	0	&	0	&	0	&	0	&	0	\\
		2008	&	0	&	0	&	0	&	0	&	0	&	0	&	0	&	0	&	0	&	0	&	0	\\
		2009	&	0	&	0	&	0	&	3	&	5	&	9	&	4	&	6	&	10	&	6	&	2	\\
		2009	&	0	&	0	&	0	&	0	&	0	&	2	&	0	&	1	&	0	&	0	&	0	\\
		2010	&	0	&	0	&	0	&	0	&	1	&	0	&	0	&	0	&	0	&	0	&	0	\\
		2011	&	0	&	0	&	0	&	0	&	1	&	3	&	3	&	4	&	2	&	1	&	0	\\
		2013	&	0	&	0	&	0	&	0	&	0	&	0	&	0	&	1	&	1	&	1	&	1	\\
		2013	&	0	&	0	&	0	&	0	&	0	&	0	&	0	&	0	&	1	&	0	&	0	\\
		2013	&	0	&	0	&	0	&	0	&	0	&	0	&	0	&	0	&	0	&	0	&	0	\\
		2013	&	0	&	0	&	0	&	0	&	0	&	0	&	0	&	0	&	0	&	0	&	0	\\
		2014	&	0	&	0	&	0	&	0	&	0	&	0	&	0	&	0	&	0	&	0	&	0	\\
		2015	&	0	&	0	&	0	&	0	&	0	&	0	&	0	&	0	&	1	&	1	&	0	\\
		2015	&	0	&	0	&	0	&	0	&	0	&	0	&	0	&	0	&	0	&	0	&	0	\\
		2015	&	0	&	0	&	0	&	0	&	0	&	0	&	0	&	0	&	0	&	0	&	0	\\
		2015	&	0	&	0	&	0	&	0	&	0	&	0	&	0	&	0	&	0	&	0	&	0	\\
		2015	&	0	&	0	&	0	&	0	&	0	&	0	&	0	&	0	&	0	&	0	&	0	\\
		2015	&	0	&	0	&	0	&	0	&	0	&	0	&	0	&	0	&	0	&	0	&	0	\\
		2016	&	0	&	0	&	0	&	0	&	0	&	0	&	0	&	0	&	0	&	1	&	0	\\
		2016	&	0	&	0	&	0	&	0	&	0	&	0	&	0	&	0	&	0	&	0	&	0	\\
		2016	&	0	&	0	&	0	&	0	&	0	&	0	&	0	&	0	&	0	&	0	&	0	\\
		2016	&	0	&	0	&	0	&	0	&	0	&	0	&	0	&	0	&	0	&	0	&	0	\\
		2017	&	0	&	0	&	0	&	0	&	0	&	0	&	0	&	0	&	0	&	0	&	0	\\
		$EM^{'}$ &2.24 &3.0 &3.32 &3.87 &3.46 &3.74 &4.47&3.74 &4 &3.32 &2.24\\		 
		\hline
	\end{tabular}
\end{table}

The elements of $EM^{'}$-index sequence are 2.24, 3.0, 3.32, 3.87, 3.46, 3.74, 4.47, 3.74, 4, 3.32 \& 2.24 and the $EM^{'}-$index sequence value is 37.40, whereas the EM-index sequence value is 29.17. Looking at these values, one can clearly see that $h$-tail articles with at least one citation also have significance in assessing scientific impact of a scholar. Figure \ref{fig:Result of $EM^{'}$-index sequence of all authors} shows the $EM^{'}$-index sequence values of all authors. Figure \ref{fig:Comparative result of h-index sequences, EM-index sequences and $EM^{'}$-index sequence of all authors} compares $EM^{'}$-index sequence value, h-index sequence value and EM-index sequence value for all authors.  \\\\\\\\\\
\begin{figure}[H]
	\centering
	\includegraphics[width=5in,height=3in,keepaspectratio]{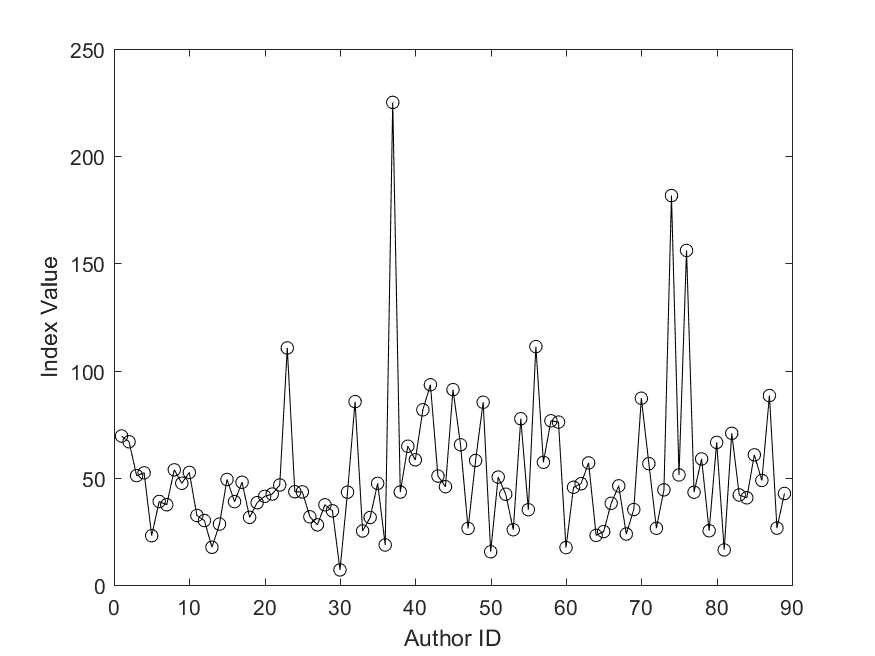}
	\caption{The $EM^{'}$-index sequence of all Scholars}
	\label{fig:Result of $EM^{'}$-index sequence of all authors}
\end{figure} 
 \begin{figure}[H]
 	\centering
 	\includegraphics[width=5in,height=4in,keepaspectratio]{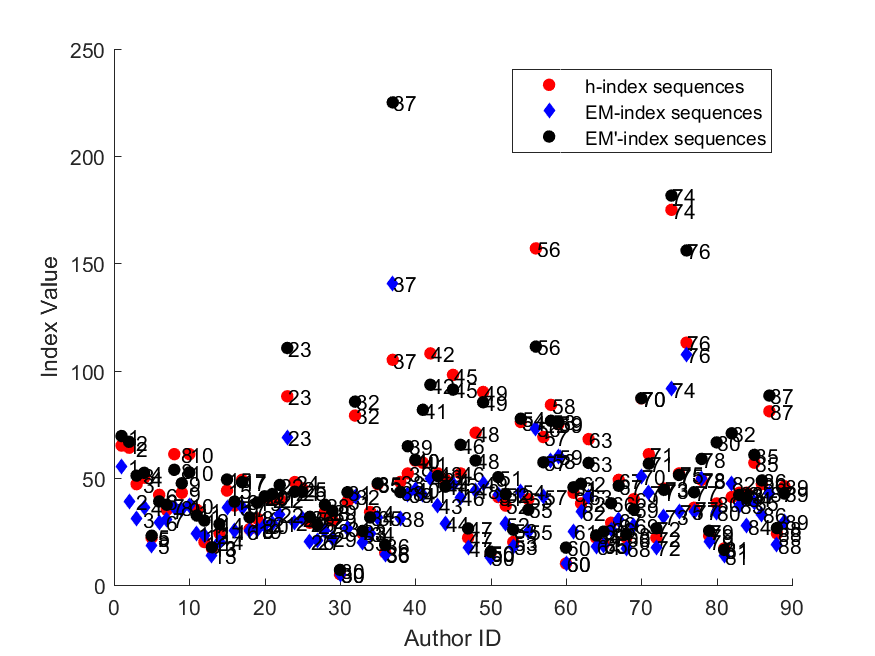}
 	\caption{Comparison of the h-index sequence, EM-index sequence and $EM^{'}$-index sequence of all Scholars}
 	\label{fig:Comparative result of h-index sequences, EM-index sequences and $EM^{'}$-index sequence of all authors}
 \end{figure} 
To demonstrate the impact of tail articles in the scientific assessment  of scholars, we performed a component level analysis on top 4 authors based on $EM^{'}$-index sequence with reference of EM-index sequence as shown in Figure \ref{fig:Comparative result of EM-index sequences and $EM^{'}$-index sequence of top 4 authors}. Looking at Figure \ref{fig:Comparative result of EM-index sequences and $EM^{'}$-index sequence of top 4 authors}, it is clear that the $EM^{'}$-index sequence gives better result than the EM-index sequence by appropriately capturing the impact of citation counts of tail articles. For example, consider Author ID-37 (see Figure \ref{fig:Comparative result of EM-index sequences and $EM^{'}$-index sequence of top 4 authors}(a)). The author's EM-index and $EM^{'}$ index sequence values are very much the same for the first 10 years -- the same is corroborated by a very low number of citations of tail articles during these years. However, after the $10^{th}$ year, the citation count of tail articles increases significantly -- making $EM^{'}$ sequence values significantly greater than  EM-index sequence values. 
\begin{figure}[H]
	\begin{subfigure}{.48\textwidth}
		\centering
		\includegraphics[width=2.8in,height=2in,keepaspectratio]{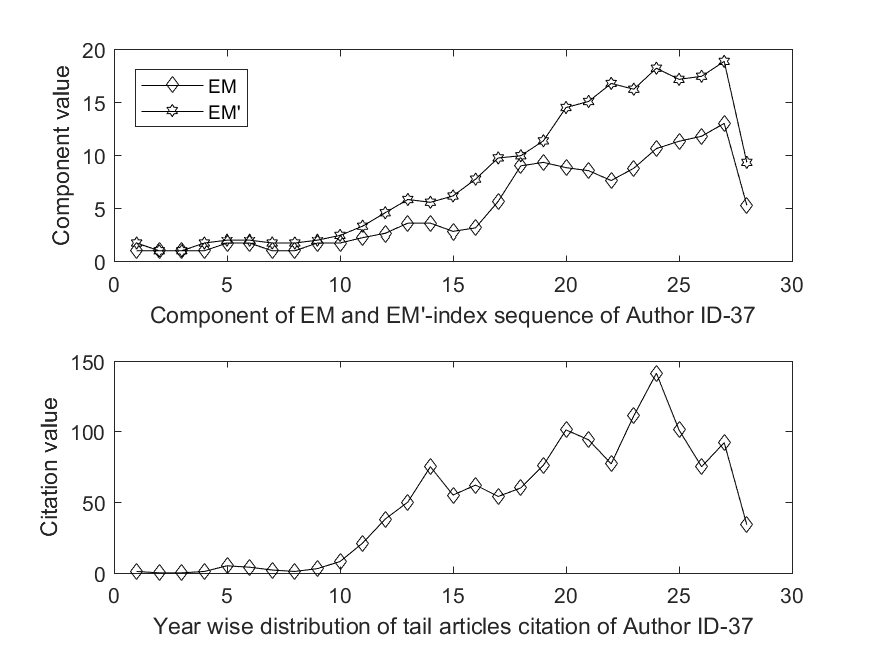}
		\caption{}
		\label{fig:a}
	\end{subfigure}
	\begin{subfigure}{.5\textwidth}
		\centering
		\includegraphics[width=2.8in,height=2in,keepaspectratio]{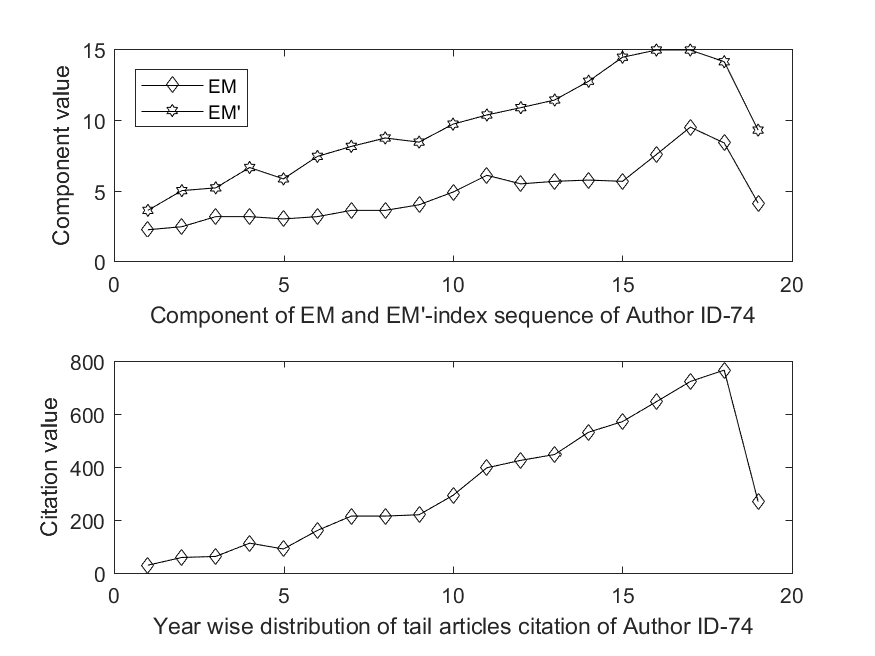}
		\caption{}
		\label{fig:b}
	\end{subfigure}
	\begin{subfigure}{.5\textwidth}
		\centering
		\includegraphics[width=2.8in,height=2in,keepaspectratio]{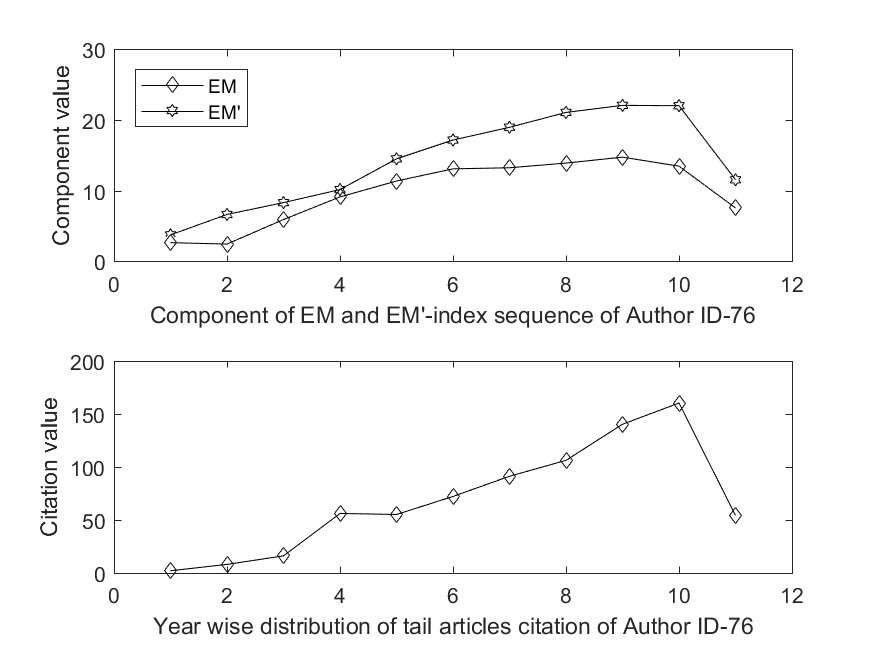}
		\caption{}
		\label{fig:b}
	\end{subfigure}
	\begin{subfigure}{.5\textwidth}
		\centering
		\includegraphics[width=2.8in,height=2in,keepaspectratio]{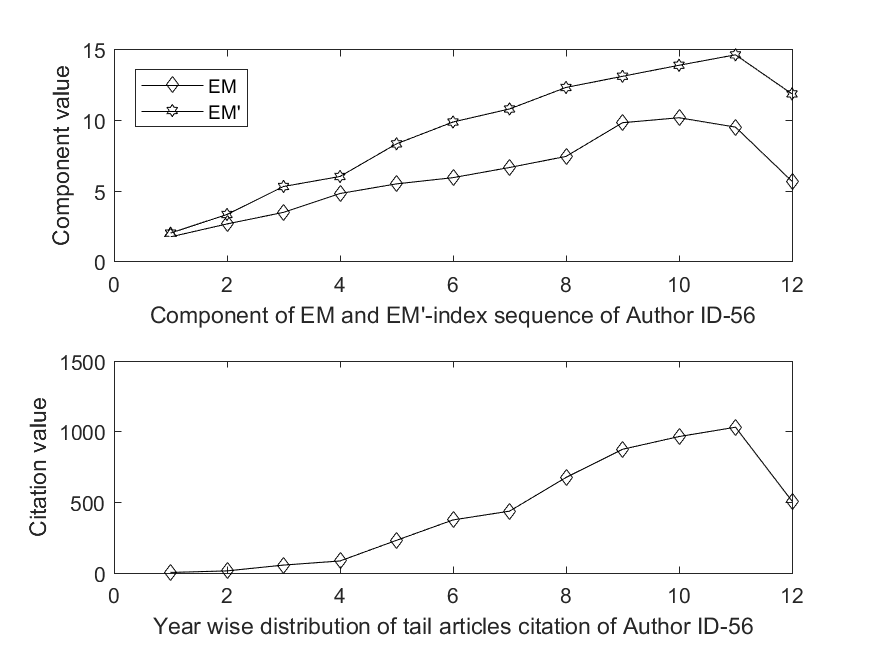}
		\caption{}
		\label{fig:b}
	\end{subfigure}
	\caption{Comparative result of EM-index sequences and $EM^{'}$-index sequence of top 4 authors with their tail articles citation distributions.}
	\label{fig:Comparative result of EM-index sequences and $EM^{'}$-index sequence of top 4 authors}
\end{figure}
 
Another extreme analysis can be seen in Figure \ref{fig:Comparative result of EM-index sequences and $EM^{'}$-index sequence of author 78 and 42}. In Figure \ref{fig:Comparative result of EM-index sequences and $EM^{'}$-index sequence of author 78 and 42}, we performed a component level analysis of authors 78 and 42 with respect to EM and $EM^{'}$ sequences. As can be seen in Figure \ref{fig:Comparative result of EM-index sequences and $EM^{'}$-index sequence of author 78 and 42}, authors 78 \& 42 have similar EM-index sequence value, however the $EM^{'}$-index sequence values exhibit significant difference. These two component level analyses show the impact of tail articles in scientific assessment of scholars and could be used as an effective alternative.

\begin{figure}[]
	\centering
	\includegraphics[width=6in,height=4in,keepaspectratio]{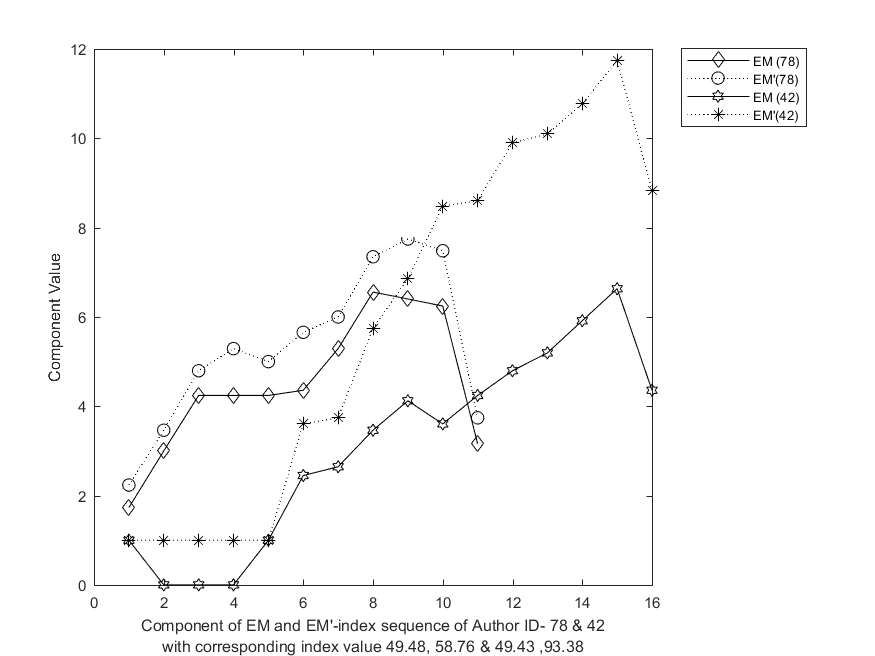}
	\caption{Component level comparison of EM-index sequence and $EM^{'}$-index sequence of Author ID- 78 \& 42 }
	\label{fig:Comparative result of EM-index sequences and $EM^{'}$-index sequence of author 78 and 42}
\end{figure} 

Table \ref{tab:A Comparative Result of h-index, EM-index and $EM^{'}$-index sequences with corresponding rank} shows a rank-based comparison based on $EM^{'}$-index sequence value, h-index sequence value and EM-index sequence value. The comparative results show that the $EM^{'}$-index sequence gives better results than others and also one can clearly see that how the excess citation and tail articles citation count affect the scientific assessment of scholars. The result of the $EM^{'}$-index sequence shows the importance of tail publications' citation count.  If we consider the impact of author id: 76, whose total citation count is 19841, h-core citation count: 12769, excess citation: 6312 and the h-tail citation is 760. The h-index sequence used only the h-core citation count and left a huge amount of citation count (7072), the EM-index sequence considers the h-core citation and excess citation count and left only few amount of citations (760), whereas the  $EM^{'}$-index sequence considers overall citation count in scientific assessment. Finally, we can conclude that the excess citation of h-core and tail articles citations gives a fair contribution in scientific assessment of scholars.  \\\\\\\\\\
\inputencoding{utf8}
\begin{table}[h!]
	\renewcommand{\arraystretch}{1.3}
	\caption{Comparison of h-index, EM-index and $EM^{'}$-index sequence with corresponding rank}
	\label{tab:A Comparative Result of h-index, EM-index and $EM^{'}$-index sequences with corresponding rank}
	\centering
	\begin{tabular}{p{1cm}p{1.5cm}p{1cm}p{1.5cm}p{2cm}p{1cm}p{1.5cm}p{2cm}p{1cm}}	\hline
		\textbf{ID}      &\textbf{h-index Sequence}&\textbf{Rank} &\textbf{Excess Citations}    &\textbf{EM-index Sequence} &\textbf{Rank} &\textbf{Tail Citations} &\textbf{$EM^{'}$-index Sequence} &\textbf{Rank} \\\hline
	1	&	65	&	18	&	897	&	55.24	&	8	&	606	&	69.47	&	17	\\
	2	&	64	&	19	&	401	&	39.06	&	27	&	1433	&	66.78	&	18	\\
	3	&	47	&	37	&	155	&	30.85	&	47	&	672	&	50.99	&	33	\\
	4	&	50	&	30	&	317	&	36.16	&	33	&	238	&	52.29	&	31	\\
	5	&	22	&	76	&	22	&	18.39	&	77	&	21	&	22.94	&	83	\\
	6	&	42	&	48	&	159	&	29.17	&	51	&	181	&	38.93	&	58	\\
	7	&	35	&	62	&	170	&	30.51	&	48	&	34	&	37.40	&	62	\\
	8	&	61	&	20	&	416	&	36.07	&	34	&	480	&	53.69	&	29	\\
	9	&	43	&	44	&	207	&	35.29	&	36	&	140	&	47.42	&	39	\\
	10	&	61	&	21	&	294	&	37.23	&	29	&	484	&	52.46	&	30	\\
	11	&	35	&	63	&	1	&	23.98	&	68	&	145	&	32.36	&	67	\\
	12	&	20	&	81	&	146	&	23.00	&	69	&	4	&	30.01	&	71	\\
	13	&	17	&	84	&	6	&	13.73	&	85	&	30	&	17.56	&	85	\\
	14	&	24	&	72	&	45	&	20.44	&	72	&	93	&	28.36	&	72	\\
	15	&	44	&	43	&	313	&	36.21	&	32	&	260	&	49.18	&	36	\\
	16	&	37	&	59	&	97	&	25.17	&	64	&	326	&	38.84	&	59	\\
	17	&	49	&	31	&	214	&	35.97	&	35	&	267	&	47.89	&	38	\\
	18	&	26	&	71	&	204	&	25.53	&	63	&	24	&	31.51	&	69	\\
	19	&	31	&	66	&	218	&	25.64	&	61	&	136	&	38.37	&	60	\\
	20	&	41	&	49	&	118	&	27.79	&	57	&	363	&	41.33	&	56	\\
		21	&	40	&	52	&	156	&	28.62	&	54	&	215	&	42.38	&	53	\\
	22	&	40	&	53	&	315	&	32.75	&	42	&	185	&	46.62	&	42	\\
		
	\hline
	\end{tabular}
\end{table}
\begin{table}[h!]
	\addtocounter{table}{-1}
	\renewcommand{\arraystretch}{1.2}
\caption{Comparison of h-index, EM-index and $EM^{'}$-index sequence with corresponding rank continued...}
\label{tab:A Comparative Result of h-index, EM-index and $EM^{'}$-index sequences with corresponding rank1}
\centering
\begin{tabular}{p{1cm}p{1.5cm}p{1cm}p{1.5cm}p{2cm}p{1cm}p{1.5cm}p{2cm}p{1cm}}	\hline
	\textbf{ID}      &\textbf{h-index Sequence}&\textbf{Rank} &\textbf{Excess Citations}    &\textbf{EM-index Sequence} &\textbf{Rank} &\textbf{Tail Citations} &\textbf{$EM^{'}$-index Sequence} &\textbf{Rank} \\\hline

	23	&	88	&	8	&	962	&	68.69	&	5	&	1121	&	110.54	&	5	\\
	24	&	48	&	35	&	121	&	29.09	&	52	&	329	&	43.46	&	47	\\
	25	&	45	&	41	&	172	&	30.98	&	46	&	245	&	43.39	&	48	\\
	26	&	29	&	68	&	61	&	20.43	&	73	&	212	&	31.76	&	68	\\
	27	&	28	&	70	&	31	&	20.50	&	71	&	110	&	28.00	&	73	\\
	28	&	34	&	64	&	100	&	25.56	&	62	&	100	&	37.37	&	63	\\
	29	&	30	&	67	&	59	&	21.60	&	70	&	194	&	34.55	&	66	\\
	30	&	5	&	89	&	6	&	5.00	&	89	&	2	&	7.00	&	89	\\
	31	&	39	&	56	&	72	&	26.38	&	59	&	403	&	43.26	&	50	\\
	32	&	79	&	12	&	415	&	41.37	&	23	&	2694	&	85.50	&	10	\\
	33	&	24	&	73	&	41	&	19.85	&	75	&	37	&	25.16	&	79	\\
	34	&	34	&	65	&	67	&	24.16	&	67	&	133	&	31.46	&	70	\\
	35	&	47	&	38	&	274	&	31.51	&	44	&	275	&	47.37	&	40	\\
	36	&	15	&	86	&	11	&	14.20	&	84	&	13	&	18.63	&	84	\\
	37	&	105	&	5	&	4122	&	140.79	&	1	&	1342	&	225.12	&	1	\\
38	&	48	&	36	&	205	&	31.10	&	45	&	235	&	43.33	&	49	\\
39	&	52	&	26	&	564	&	42.25	&	21	&	201	&	64.69	&	21	\\
40	&	58	&	23	&	442	&	44.69	&	15	&	393	&	58.34	&	24	\\
41	&	57	&	24	&	943	&	42.13	&	22	&	155	&	81.70	&	12	\\
42	&	108	&	4	&	584	&	49.43	&	11	&	2724	&	93.38	&	6	\\
43	&	52	&	27	&	354	&	37.21	&	30	&	292	&	50.79	&	34	\\
44	&	49	&	32	&	134	&	28.71	&	53	&	438	&	45.82	&	44	\\
45	&	98	&	6	&	575	&	46.97	&	14	&	3037	&	91.10	&	7	\\
46	&	52	&	28	&	535	&	40.81	&	26	&	114	&	65.38	&	20	\\
47	&	22	&	77	&	49	&	17.26	&	82	&	127	&	26.34	&	76	\\
48	&	71	&	15	&	591	&	44.00	&	16	&	699	&	58.04	&	25	\\
49	&	90	&	7	&	508	&	47.42	&	13	&	2216	&	85.22	&	11	\\
50	&	15	&	87	&	32	&	12.70	&	87	&	10	&	15.46	&	88	\\
51	&	41	&	50	&	435	&	42.35	&	20	&	221	&	50.29	&	35	\\
52	&	37	&	60	&	192	&	28.49	&	55	&	211	&	42.31	&	54	\\
53	&	20	&	82	&	81	&	17.89	&	80	&	27	&	25.73	&	77	\\
54	&	76	&	13	&	440	&	43.74	&	17	&	1491	&	77.54	&	13	\\
55	&	40	&	54	&	98	&	25.01	&	66	&	177	&	35.08	&	65	\\
56	&	157	&	2	&	1930	&	73.17	&	4	&	5242	&	111.18	&	4	\\
57	&	69	&	16	&	466	&	41.33	&	25	&	586	&	57.22	&	26	\\
58	&	84	&	10	&	947	&	58.17	&	7	&	1004	&	76.57	&	14	\\
59	&	75	&	14	&	980	&	59.89	&	6	&	906	&	75.97	&	15	\\
60	&	10	&	88	&	12	&	10.00	&	88	&	30	&	17.39	&	86	\\
61	&	43	&	45	&	62	&	25.07	&	65	&	470	&	45.55	&	45	\\
62	&	38	&	57	&	625	&	34.08	&	39	&	300	&	47.21	&	41	\\
63	&	68	&	17	&	552	&	41.35	&	24	&	564	&	56.92	&	27	\\
64	&	22	&	78	&	41	&	17.99	&	79	&	46	&	23.09	&	82	\\
65	&	19	&	83	&	62	&	18.39	&	78	&	42	&	24.87	&	80	\\
66	&	29	&	69	&	237	&	25.96	&	60	&	74	&	38.07	&	61	\\
67	&	49	&	33	&	95	&	30.36	&	49	&	345	&	46.21	&	43	\\
68	&	22	&	79	&	34	&	17.02	&	83	&	66	&	23.73	&	81	\\
69	&	40	&	55	&	98	&	27.83	&	56	&	189	&	35.15	&	64	\\
70	&	87	&	9	&	478	&	50.67	&	9	&	2152	&	87.09	&	9	\\
71	&	61	&	22	&	633	&	43.30	&	18	&	343	&	56.56	&	28	\\
72	&	22	&	80	&	53	&	17.39	&	81	&	38	&	26.44	&	75	\\
73	&	45	&	42	&	226	&	31.87	&	43	&	250	&	44.36	&	46	\\
74	&	175	&	1	&	1774	&	91.43	&	3	&	6215	&	181.65	&	2	\\
75	&	52	&	29	&	340	&	34.46	&	37	&	316	&	51.30	&	32	\\

	\hline
\end{tabular}
\end{table}
\begin{table}[h!]
\addtocounter{table}{-1}
\renewcommand{\arraystretch}{1.2}
\caption{Comparison of h-index, EM-index and $EM^{'}$-index sequence with corresponding rank continued...}
\label{tab:A Comparative Result of h-index, EM-index and $EM^{'}$-index sequences with corresponding rank1}
\centering
\begin{tabular}{p{1cm}p{1.5cm}p{1cm}p{1.5cm}p{2cm}p{1cm}p{1.5cm}p{2cm}p{1cm}}	\hline
\textbf{ID}      &\textbf{h-index Sequence}&\textbf{Rank} &\textbf{Excess Citations}    &\textbf{EM-index Sequence} &\textbf{Rank} &\textbf{Tail Citations} &\textbf{$EM^{'}$-index Sequence} &\textbf{Rank} \\\hline
76	&	113	&	3	&	6312	&	107.41	&	2	&	760	&	156.04	&	3	\\
77	&	36	&	61	&	220	&	34.26	&	38	&	126	&	43.23	&	51	\\

78	&	49	&	34	&	656	&	49.48	&	10	&	432	&	58.76	&	23	\\
79	&	23	&	75	&	44	&	20.16	&	74	&	19	&	25.25	&	78	\\
80	&	38	&	58	&	631	&	33.80	&	40	&	51	&	66.48	&	19	\\
81	&	17	&	85	&	25	&	13.49	&	86	&	6	&	16.33	&	87	\\
82	&	41	&	51	&	779	&	47.50	&	12	&	520	&	70.73	&	16	\\
83	&	43	&	46	&	336	&	37.21	&	31	&	60	&	41.94	&	55	\\
84	&	43	&	47	&	108	&	27.58	&	58	&	314	&	40.63	&	57	\\
85	&	57	&	25	&	298	&	38.37	&	28	&	655	&	60.68	&	22	\\
86	&	47	&	39	&	209	&	33.06	&	41	&	402	&	48.75	&	37	\\
87	&	81	&	11	&	439	&	42.58	&	19	&	2819	&	88.33	&	8	\\
88	&	24	&	74	&	29	&	18.65	&	76	&	70	&	26.50	&	74	\\
89	&	46	&	40	&	94	&	29.57	&	50	&	237	&	42.65	&	52	\\
		\hline
	\end{tabular}
\end{table}

To find the correlation between indices, the Spearman Rank Correlation test has been performed on h-index sequence value, EM-index sequence value and $EM^{'}$-index sequence value.  Table \ref{tab:Result of Spearman rank correlation between h-index sequences, EM-index sequences and $EM^{'}$-index sequences} shows the result of correlation test between indices. The result shows that  all  the indices are highly correlated mutually. Hence, it can be said that the proposed EM and $EM^{'}$ index sequences  are on the same path as the already established indicators.

\begin{table}[hbt]
	\renewcommand{\arraystretch}{1.2}
	\caption{Result of Spearman rank correlation between h-index sequence, EM-index sequence and $EM^{'}$-index sequence}
	\label{tab:Result of Spearman rank correlation between h-index sequences, EM-index sequences and $EM^{'}$-index sequences}
	\centering
	\begin{tabular}{p{3cm}p{3cm}p{3cm}p{3.2cm}}		\hline
		\textbf{}	&	\textbf{h-index sequence}	&	\textbf{EM-index sequence}	&	\textbf{$EM^{'}$-index sequence}\\\hline
		h-index sequence & 1 &0.93 &0.94 \\
		EM-index sequence & 0.93 &1 &0.96\\
		$EM^{'}$-index sequence &0.94  &0.96 &1\\
			 
		\hline
	\end{tabular}
\end{table}

\section{Conclusion}
The h-index sequence uses the series of h-index based on individual year citation count along with the career-span of research for scientific assessment of scholars. A set of indices values helps users to discriminate between the performance of scholars at a particular stage of their careers or their whole careers. However, the h-index sequence completely ignored the importance of excess citations. In this article, the EM-index sequence and $EM^{'}$-index sequence has been discussed. The EM-index and $EM^{'}$-index sequence values can provide a superior alternative -- when compared to $h$-index sequence values --  for assessing the scientific impact of scholars. 

\begin{acknowledgements}
	The authors would like to acknowledge the help of Ministry of Electronics \& Information Technology (MeitY),Government of India for supporting the financial assistance during research work through “Visvesvaraya PhD Scheme for Electronics \& IT”.
\end{acknowledgements}

\bibliographystyle{apalike}      
\bibliography{reportnew}                

\end{document}